\documentclass{article}
\usepackage{amssymb}
\usepackage{epsfig}
\usepackage{makeidx}
\usepackage{graphicx}
\usepackage{multicol}
\usepackage{amsmath}
\usepackage{amsfonts}

\begin{document}

\title{\textbf{The method of dynamic projection operators in the theory of hyperbolic systems of partial differential equations with variable coefficients}}
\author{S.Leble$^{1,2}$, I.Vereshchagina$^{2}$ \\ \small 1. Gdansk University of Technology,
Poland, \\ \small 2. I. Kant BFU , Kaliningrad,  Russia.}

\maketitle
\begin{abstract}

We consider a generalization of the  projecting operators method for the case of Cauchy problem for systems of 1D evolution differential equations of first order with variable coefficients. It is supposed that the coefficients dependence on the only variable x  is weak, that is described by a small parameter introduction. Such problem corresponds, for example, to the case of wave propagation in a weakly inhomogeneous medium.  As an example, we specify the problem to adiabatic acoustics. 
For the  Cauchy problem, to fix unidirectional modes,  the projection operators are constructed. The method of successive approximations (perturbation theory) is developed and based on pseudodifferential operators theory. The application of these projection operators allows to obtain approximate evolution equations corresponding to the separated  directed waves. 
\end{abstract}

\normalsize

\section{Introduction}

The main idea of a division of  solutions space of an evolution equation, to subspaces that correspond to so-called   dispersion relations  (roots of dispersion equation that link  frequency and wave vector), go up to the paper of Chu and Kovasznay \cite{CK}. The wave vector and therefore - frequency are introduced via Fourier transformation in space coordinates that is effective almost exclusively in the case of homogeneous background state (coefficients of the equations independent on the coordinates). Links between dynamical variables  are specified for such subspaces. A development of this idea is to combine the equations of the system under investigation in such manner that allows to "diagonalize" the evolution operator. Technically, both actions may be realized via application of a projecting procedure \cite{Ver,L}. More precisely, we use the idempotents built on eigenvectors of the evolution operator \cite{BKL, L, Per1}. The projectors solve both tasks: combine the equations and change the dynamical variables.  The idea of a projecting in similar approach later was also formulated in \cite{Pie}. Nonlinearity presence in a problem within such  approach was realized in spirit of perturbation theory: the nonlinear terms were combined by the same (built in linearized theory) projecting operators \cite{L,P0,P3}, doing the step which have not been made in \cite{CK}. The only example of nonlinear non-perturbative corrections account was realized by Riemann and, in the projecting technique content,  in Perelomova's \cite{P}. The general problems following Riemann results (Riemann waves) were investigated in publications of Z. Peradzinski \cite{Pi}.

Briefly, the \textit{idea of this approach} may be described by a simple example of constant coefficients as follows.  Consider an evolution problem as the system of two equations with constant coefficients.
\begin{equation}\label{evx}
\frac{\partial u(x,t)}{\partial t }
-a\frac{\partial u(x,t)}{\partial x}-b\frac{\partial v(x,t)}{\partial x}=0,
\end{equation}
\begin{equation}\label{evx1}
\frac{\partial v(x,t)}{\partial t }
-c\frac{\partial u(x,t)}{\partial x}-d\frac{\partial v(x,t)}{\partial x}=0,
\end{equation}
A compact  matrix form of (\ref{evx},\ref{evx1})  
\begin{equation}\label{ev}
	\psi_t=L\psi,
\end{equation}
with
\begin{equation}\label{mf}
\psi=\left(
          \begin{array}{c}
              u \\
               v \\
          \end{array}
          \right)\textrm{ and } L=\left(
          \begin{array}{cc}
               a\partial_{x} & b\partial_{x}\\
               c\partial_{x} & d\partial_{x}\\
          \end{array}
          \right),
\end{equation}
introduces the evolution operator $L$ and a state $\psi$ of a system.
Fourier transformation in x
\begin{equation}\label{eq:D_relation_to_E}
  u(x)=\frac{1}{\sqrt{2\pi}}\int_{-\infty} ^{\infty}\tilde{u}(k)e^{ikx}dk,\, v(x)=\frac{1}{\sqrt{2\pi}}\int_{-\infty} ^{\infty}\tilde{v}(k)e^{ikx}dk,
\end{equation}
may be written as matrix substitution
\begin{equation}
	\psi=F\tilde{\psi}.
\end{equation}
Hence, in compact notation of derivatives by index, it yields ordinary differential equations system 
\begin{equation}\label{evk}
	\tilde{\psi_t}=F^{-1}LF\tilde{\psi}=\tilde{L}\tilde{\psi},
\end{equation}
where, the k-representation of the evolution operator is 
\begin{equation}\label{Lm}
\tilde{L}=ik\left(
          \begin{array}{cc}
               a  & b\\
               c & d\\
          \end{array}
          \right).
\end{equation}

\textbf{ General note.} A matrix $n\times n$ eigenvalue problem
\begin{equation}
	\tilde{L}\phi=\lambda\phi
\end{equation}
introduces two subspaces, which we   represent by the matrix of solutions $\Psi$
\begin{equation}\label{ef}
	\tilde{L}\Psi=\Psi\Lambda,
\end{equation}
with diagonal matrix $\Lambda=diag\{\lambda_1,...,\lambda_n\}$. We would choose the normalization of the eigenvectors such that the first component is unit.
Suppose, the inverse matrix $\Psi^{-1}$ exists and
\begin{equation}\label{Psi1}
\Psi^{-1}	\tilde{L} =\Lambda\Psi^{-1}.
\end{equation}
Multiplying from the left side by $\Psi^{-1}$ gives
\begin{equation}
\tilde L = \Psi \Lambda \Psi^ {-1},
 \end{equation}
or, in components, it give the spectral decomposition of the matrix $\tilde L$
\begin{equation}\label{p}
\tilde L_{ij} = \Psi_{ik} \Lambda _{kl}\Psi^ {-1}_{lj}=
\Psi_{ik} \lambda_k\Psi^ {-1}_{kj} =
\sum_k \lambda_{k}\Psi_{ik} \Psi^ {-1}_{kj}=
\sum_k  \lambda_{k} (\tilde P_k)_{ij}.
 \end{equation}

Returning to the $2\otimes 2$ example, let us search for a matrix $\tilde P_{i}, \, i=1.2$, such that $\tilde P_{i} \Psi = \Psi_{i}$   to be eigenvectors of the evolution matrix in Eq. (\ref{evk}). Moreover, the standard properties of orthogonal projecting operators
\begin{equation}\label{projcond}
\tilde P_{i}*\tilde P_{j} = 0, \quad \tilde P^{2}_{i} = \tilde P_{i}, \quad \sum_{i}\tilde P_{i} = 1
\end{equation}
are implied.
 Really,    if $\Psi^{-1}$ exists 
 we can apply the results from the general note  \eqref{p}, or one can prove \cite{LZ} that
\begin{equation}\label{proj}
\tilde P_{i}  = \Psi_i\otimes\Psi^{-1}_i,  
\end{equation}
where $\Psi_i$ is the i-th column and $\Psi^{-1}_i$ - i-th row of the corresponding inverse matrix  and the identity
\begin{equation}
	\tilde{L} \tilde P_i=\tilde P_i\tilde{L}
\end{equation}
 holds. 
 Explicit form of the operators and variables in the mentioned normalization is given by
\begin{equation}\label{Psij}
	\Psi=
 \left(
          \begin{array}{cc}
               1  & 1\\
              \tilde v_1 &\tilde v_2\\
          \end{array}
          \right).
\end{equation}
Then, the eigenvalues of $\tilde L$
\begin{equation}
	\lambda_{1,2}=\frac{ik}{2}[(a+d)\pm\sqrt{(a-d)^2+4bc}].
\end{equation}
The values $v_i$ are found from (\ref{ef}) 
\begin{equation}
\tilde	v_i=-\frac{i \lambda_i + ak}{bk}=\frac{(a+d)\pm\sqrt{\Delta}}{2b}-\frac{ a}{b},
\end{equation}
if $\Delta =(a-d)^2+4bc>0$, the $\frac{\lambda_i}{i}$ are real and the equations \eqref{evx} are hyperbolic, that corresponds wave propagation, because $\Pi$ is right and $\Lambda$ is left waves. Really,    $\lambda_1\neq\lambda_2,\,\exists \Psi^{-1}$ 
and we can apply the results from general note  \eqref{p}
\begin{equation}\label{P}
\tilde	P_1=\frac{1}{\tilde v_2-\tilde v_1}\left(
          \begin{array}{c}
              1 \\
              \tilde v_1 \\
          \end{array}\right)\otimes\left(\tilde v_2,-1\right),
\end{equation}
and
\begin{equation}
\tilde	P_2=\frac{1}{\tilde v_2-\tilde v_1}\left(
          \begin{array}{c}
              1 \\
               \tilde v_2 \\
          \end{array}\right)\otimes\left(-\tilde v_1,1\right).
\end{equation}

 Its x-representation counterpart $L$ and  
 the inverse Fourier transforms of $\tilde P_i$
\begin{equation}
	F\tilde P_iF^{-1} \equiv P_i
\end{equation}
 also commute; the proof is in the next line 
\begin{equation}\label{com}
	F\tilde{L}F^{-1}F \tilde P_iF^{-1}=F\tilde P_iF^{-1}F \tilde{L}F^{-1}.
\end{equation}
Projecting the evolution (\ref{evk})  gives two independent equations, whose are read as the first lines of
\begin{equation}\label{eq}
	(P_i\psi)_t=LP_i\psi,
\end{equation}
the commutation (\ref{com}) is taken into account. In new variables 
\begin{equation}
	\Pi=(P_1\psi)_1,\,\Lambda=(P_2\psi)_1.
\end{equation}
whence
\begin{equation}\label{Px}
	P_1=\frac{1}{v_2-v_1}\left(
          \begin{array}{c}
              1 \\
               v_1 \\
          \end{array}\right)\otimes\left(v_2,-1\right),
\end{equation}
and
\begin{equation}
	P_2=\frac{1}{v_2-v_1}\left(
          \begin{array}{c}
              1 \\
               v_2 \\
          \end{array}\right)\otimes\left(-v_1,1\right).
\end{equation}
are constant matrices, because $v_{1,2}= \frac{a+d\pm\sqrt{\Delta}}{2b}-\frac{a}{b}$.

Applying the operator $P_1$ to the vector $\psi$ yields
\begin{equation}\label{Pi}
	\Pi=(P_1\psi)_1=\frac{1}{v_2-v_1}(v_2 u-v),
	\end{equation}
	\begin{equation}
	\Lambda=(P_2\psi)_1= \frac{1}{v_2-v_1}(- v_1 u + v).
\end{equation}
It splits the original system to  the system of independent equations directly from (\ref{eq}).
\begin{equation}\label{Pi}
	\Pi_t=(a+bv_1) \partial_{x}\Pi,\,\Lambda_t=(a+bv_2)\partial_{x}\Lambda.
\end{equation}
Such system naturally describes unidirectional one-dimensional waves propagation correspondent to initial problem, defined by  the Cauchy problem that has elegant formulation in the context of projecting procedure
\begin{equation}\label{Pi2}
	\Pi(x,0)=(P_1\psi(x,0))_1,\,\Lambda(x,0)=(P_2\psi(x,0))_1.
\end{equation}

Let us take the typical for physical applications case $a=d=0$, equivalent to classic string equation (see also Sec.6). Now $\lambda_{1,2}=\pm ik\sqrt{bc},\,v_{1,2}=\pm  \sqrt{\frac{c}{b}}$, therefore (\ref{Px}) significantly
simplifies
\begin{equation}\label{Px1}
	P_1=\frac{1}{ -2\sqrt{\frac{c}{b}}}\left(
          \begin{array}{c}
              1, 
               \sqrt{\frac{c}{b}} \\
          \end{array}\right)\otimes\left( -\sqrt{\frac{c}{b}},-1\right)=\frac{1}{ 2}\left(
          \begin{array}{cc}
              1 & \sqrt{\frac{b}{c}}\\
               \sqrt{\frac{c}{b}}&1 \\
          \end{array}\right),
\end{equation}
and yields 
\begin{equation}\label{Pi3}
	\Pi_t=\sqrt{bc} \partial_{x}\Pi,\,\Lambda_t=-\sqrt{bc} \partial_{x}\Lambda.
\end{equation}
The evolution corresponds opposite one-dimensional 
   acoustic  \cite{P} or electromagnetic waves \cite{KL} and many others, for example,  for the electromagnetic environment with the division on the right and left wave in \cite{K}.
 D'Alembert formula follows directly from (\ref{Pi2}, \ref{Pi3}) and completeness identity
\begin{equation}
	\psi=(P_1+P_2)\psi.
\end{equation}

The formalism obliously is not restricted by the case of two by two matrices and dispersionless and non-dissipative case of wave theory. it is effectively applied up to 5x5 evolution operator of hydrodynamics \cite{L} and recently to 4x4 electrodynamics \cite{L,LeKupol}. In papers \cite{P,P0,Per1,Pie} it is developed for a acoustics problems and in \cite{LR} for plasma physics. There are very interesting phenomena as heating and streaming that appear in interaction of acoustic and zero frequency modes \cite{P,P3}. Such theory may be considered as a development of Heaviside operator method similarly to general operator method outlined in \cite{M}.

The challenge  of a  further development of the projecting operator method relates to problems of evolution via differential operators with coefficients dependent on coordinates \cite{M}. Its main obstacle is in eventual simplifications after Fourier transformations which cannot be directly applied now. The only example that was successfully solved relates to exponential stratification. The projecting operator in this case has matrix elements which are integral operators with kernels defined via Hankel functions \cite{P2}.

To analyze the solutions of the equations is useful to know in what physical conditions the equation was obtained namely
what is taken into account in the derivation of this equation. Traditionally, the wave fields are divided
into components   (entropic, acoustic, and vortical modes of \cite{CK}, last ones are subdivided into the right and the left waves).
The eigenvectors of the linearized system are used to build the projection operators. The projection operators
can select the relevant wave in arbitrary time. For a general inhomogeneous medium, neither Fourier transformation nor dispersion relation can be effectively.

The aim  of this investigation is based on the ideas of projecting method, but it does not rely upon the Fourier transform. We, however  use its spirit in a form of pseudifferential operators and corresponding expansion. More exactly we write directly the evolution operators  (in Fourier transform it is a parameter $\omega$ - frequency). In a solution subspace such operator is presented by a power series of operator of the derivative in basic space variable. Correspondingly the matrix elements of the projecting operators are built as similar expansions - whence are also pseudo differential operators.

\section{System of two equations with variable coefficients.}

Consider one-dimensional system of two equations  with variable coefficients. Let it be a  hyperbolic  differential equation with variable coefficients in the domain, described in introduction. 
\begin{equation}\label{dal1}
\frac{\partial u(x,t)}{\partial t }
-a(x)\frac{\partial u(x,t)}{\partial x}-b(x)\frac{\partial v(x,t)}{\partial x}=0,
\end{equation}
\begin{equation}\label{dal2}
\frac{\partial v(x,t)}{\partial t }
-c(x)\frac{\partial u(x,t)}{\partial x}-d(x)\frac{\partial v(x,t)}{\partial x}=0,
\end{equation}
where  $a,b,c,d$ are coefficients of the evolution operator. The Cauchy problem for \eqref{dal1} is specified by
\begin{equation}\label{dal3}
u(x,0)=\phi(x),\quad v(x,0)=\psi(x).
\end{equation}

Applying formal operator notations 
	
	$$
	\hat a(x)  = a(x)D,\, 
	  \hat b(x)= b(x)D, \,
	  \hat c(x) = c(x)D,\,
	  \hat d(x) = d(x)D,
$$
we write a system: 
\begin{equation}\label{f}
\hat a(x)   \tilde{u}(x) + \hat b(x)  \tilde{v}(x)= \lambda (D)\tilde{u},
\end{equation}
\begin{equation}\label{dal20}
 \hat c(x)   \tilde{u}(x) + \hat d(x)  \tilde{v}(x)= \lambda (D) \tilde{v},
\end{equation}
that define a pseudodifferential operator $\lambda (D)$.


Solving the system formally, $b\neq 0$, we start with
\begin{equation}\label{v}
   \tilde{v}(x,t)= + \hat b(x)^{-1}(\lambda -\hat a(x)   )\tilde{u}.
\end{equation}
 This relation (\ref{v}) may be considered as the link, that define the eigenvectors 
 \begin{equation}\label{ph}
\phi=\left( 
\begin{array}{c} 
	\tilde{u}\\
	\tilde{v}
	 \end{array} \right)
\end{equation}
 for each $\lambda(D)$. Plugging the link \eqref{v} into (\ref{dal20}) one obtains
\begin{equation}\label{g}
\hat c(x)   \tilde{u}(x) + (\hat d(x)  -\lambda) [\hat b(x)^{-1}(\lambda -\hat a(x)   )\tilde{u}]= 0.
\end{equation}
It gives  an  equation for the unknown operator $\lambda$ :
\begin{equation}\label{phi}
\{   -\lambda \hat b(x)^{-1} \lambda + \lambda\hat b(x)^{-1}\hat a(x)+ \hat d(x)\hat b(x)^{-1}\lambda -\hat d(x)\hat b(x)^{-1}\hat a(x)+\hat c(x) \}  \tilde{u}(x)= 0.
\end{equation}

\section{Expansions and approximation}

Suppose the operator $\lambda(D)$ is generally a pseudo-differential one 
\begin{equation}\label{lambda}
	\lambda_i(D)=\sum_{n=0}^{\infty} s_n^{(i)}(x) D^n
\end{equation}
   Plugging \eqref{lambda} into  \eqref{phi} results in
\begin{equation}\label{phi1}
\begin{array}{c}
\{   -(\sum_{n,m=-k}^{\infty} s_m^{(i)}(x) D^m)D^{-1}  b ^{-1} ( s_n^{(i)}(x) D^n) +\\
 (\sum_{n=-k}^{\infty} s_n^{(i)}(x) D^n) D^{-1}  b ^{-1}  aD+\\   d b ^{-1} D^{-1}  b ^{-1} (\sum_{n=-k}^{\infty} s_n^{(i)}(x) D^n) - d  b ^{-1}aD+cD \}  \tilde{u}(x)= 0.
 \end{array}
\end{equation}
for each mode, defined on a space S, such that the series \eqref{lambda} converges \cite{ps}, having in mind a possibility to use a finite number of terms on a subspace $S_)\in S$. It corresponds to so-called long wave approximation in many works devotes to wave propagation theory \cite{L}.

Restricting ourselves by the three-term approximation $\lambda = p+qD+rD^{2}$, 
we get:
\begin{equation}\label{phi2}
\begin{array}{c}
\{   -(p+qD+rD^{2}) D^{-1}  b ^{-1} (p+qD+rD^{2}) + (p+qD+rD^{2}) D^{-1}  b ^{-1}  aD+\\   d b ^{-1} (p+qD+rD^{2}) - d  b ^{-1}aD+cD \}  \tilde{u}(x)= 0.
\end{array}
\end{equation}
Next, suppose that   derivatives of the original system   \eqref{dal1} coefficients are of the minor order compared with the coefficients itself. Then, 
equalizing the coefficients by powers of D, taking the order of derivatives into account,
\begin{equation}
\begin{array}{c}
D^{0}:    -p b ^{-1} q + p(b ^{-1}r)'-q b ^{-1} p+ p b ^{-1}a-r(b ^{-1}p)'+d b ^{-1}a=0 \\
D^{1}:    -pb ^{-1}r -q b ^{-1} q + r b ^{-1}p -r(b ^{-1}q)'+ q b ^{-1} a + r(b ^{-1}a)' + d b ^{-1}q+c-d b ^{-1}q=0\\
D^{2}:    -q b ^{-1} r - r b ^{-1} q - r(b ^{-1}r)'+ r b ^{-1} a + d b ^{-1} r = 0.  \\
 \end{array}	
\end{equation}
The commutation relations are used in transformation.
 
Multiplying the second equation on $b\neq 0$ will get:
\begin{equation}\label{qs}
\begin{array}{c}
	p=0,\, r=0\\
	q_{\pm}=\frac{(a+d)\pm \sqrt{(a-d)^2+4bc}}{2}
	 \end{array}	
\end{equation}
The relation coincide with ones from introduction in the order under consideration 
\begin{equation}\label{D}
	\lambda_{\pm}=q_{\pm}D,
\end{equation}
  that support the result.

\section{Projecting operators}

Going to generalization of the method described in introduction let us
consider a 2x2 matrix with operator-valued elements;
\begin{equation}\label{P}
   P=\left(
       \begin{array}{cc}
         p & \pi \\
         \xi & \eta \\
       \end{array}
     \right)
\end{equation}
with basic determining idempotent condition
$$
P^2=P.
$$

It immediately yields
\begin{equation}\label{P}
   P=\left(
       \begin{array}{cc}
         p & \pi \\
       \pi^{-1} (p-p^2) & 1-\pi^{-1}p\pi \\
       \end{array}.
     \right)
\end{equation}
There are possibilities of operators $p,\pi$ choice, that fix the projection subspaces \eqref{ph}
\begin{equation}\label{P}
   \left(
       \begin{array}{cc}
         p & \pi \\
       \pi^{-1} (p-p^2) & 1-\pi^{-1}p\pi \\
       \end{array}
     \right)
		\left(
		\begin{array}{c}
		u\\ v
		\end{array}\right)
		=\left(\begin{array}{c}
		\widetilde{u} \\\widetilde{v}
		\end{array}\right).
\end{equation}
Using this equality and the condition of completeness $P_++P_-=I$, we obtain the explicit form of the projection operators, that correspond to two versions of $\lambda$ for both $q_{\pm}$ given by \eqref{qs}.

\section{Particular case}

We consider a  more compact, still hyperbolic case  $a=0, d=0,\,bc>0 $:
\begin{equation}\label{1}
\frac{\partial u(x,t)}{\partial t }
-b(x)\frac{\partial v(x,t)}{\partial x}=0,
\end{equation}
\begin{equation}\label{2}
\frac{\partial v(x,t)}{\partial t }
-c(x)\frac{\partial u(x,t)}{\partial x}=0,
\end{equation}
Projecting operators in this case are calculated via the definition \eqref{P} for the projection subspaces \eqref{ph}  :
\begin{equation}
  P_{1,2}=\frac{1}{2} \left(
       \begin{array}{cc}
         1 & \pm (f-D^{-1}f^{'})^{-1} \\
         \pm(f-D^{-1}f^{'}) & 1\\\
       \end{array}
     \right),
\end{equation}
where $f= \sqrt\frac{c(x)}{b(x)}.$
Now the evolution 
operator $L$ (see, for example, \eqref{mf}), in the same notations, simplifies:
\begin{equation}
  L=\left(
       \begin{array}{cc}
         0 & b(x)D\\
         \ c(x)D& 0 
       \end{array}
     \right).
\end{equation}

The commutator $L$ and $P_1$ is equal to
\begin{equation}
  [P_{1}, L]=\left(
        \begin{array}{cc}
          D^{-1}f^{-1}D cD- bfD & 0 \\
					         0 &  D^{-1}fD bD - cf^{-1}D 
       \end{array}
     \right),
     \end{equation}
because the identities $f-D^{-1}f^{'}=D^{-1}fD$ and $(f-D^{-1}f^{'})^{-1}=D^{-1}f^{-1}D$ hold.

Condition that the commutator is zero can be written as
\begin{equation}
D^{-1}f^{'} b f=0,
\end{equation}
 or with the expression for f:
\begin{equation}
\frac{1}{2} D^{-1} (c^{'} -\frac{b^{'}}{b} c) = 0.
\end{equation}
 It fix the case of complete reduction (diagonalisation) of evolution operator.
 
 As the further development of the method we suggest an approximate procedure (see e.g. \cite{PL}) 
Using the  projecting operators  we shall found a new equations for left and right waves, splitting the problem of evolution. The approximate splitting is achieved if one could neglect the commutators of $P_{1,2}$ and $L$. It is possible if the coefficients $b,c$ are of the zero order ($\cong O(1)$), while the order of the  derivative $(\frac{c}{b})'$ is of a higher order, e.g. $\cong O(\epsilon)$. Acting by $P_1$ to the system \eqref{dal1} 
 \begin{equation}
\begin{array}{cc}
  P_{1,2}\Psi_t=P_{1,2}L\Psi, 
	\end{array}
	\end{equation}
or, approximately	
	\begin{equation}
	\begin{array}{cc}
  (P_{1,2}\Psi)_{t}=L(P_{1,2}\Psi).  
	 \end{array}
	\end{equation}

	Where 
	\begin{equation}
  P_{1}\Psi=\frac{1}{2} \left(
       \begin{array}{cc}
         1 &  D^{-1}f^{-1}D \\
         D^{-1}fD & 1\\\
       \end{array}
     \right)
			\left(
		\begin{array}{c}
		u\\ v
		\end{array}\right)
		=\frac{1}{2} \left(
       \begin{array}{cc}
         \Pi  \\
         (f-D^{-1}f^{'}) \Pi\\
       \end{array}
     \right)
\end{equation}

		\begin{equation}
  P_{2}\Psi=\frac{1}{2} \left(
       \begin{array}{cc}
         1 &  -D^{-1}f^{-1}D \\
         -D^{-1}fD & 1\\\
       \end{array}
     \right)
			\left(
		\begin{array}{c}
		u\\ v
		\end{array}\right)
		=\frac{1}{2} \left(
       \begin{array}{cc}
         \Lambda  \\
         -D^{-1}fD \Lambda \\
       \end{array}
     \right)
\end{equation}
reading the first lines of the relations yields	
\begin{equation}\label{pi}
	  \Pi=\frac{1}{2} (u + D^{-1}f^{-1}D v),\\
			\end{equation}
and		
	\begin{equation}\label{la}
	  \Lambda=\frac{1}{2} (u - D^{-1}f^{-1}D v).\\
			\end{equation}.
			From this equations follow:
			\begin{equation}\label{uv}
	\begin{array}{cc}
  u = \Pi+\Lambda,\\
	v = (f-D^{-1}f^{'}) (\Pi-\Lambda)
	 \end{array}
	\end{equation}
This relation allows to state the Cauchy problems for directed waves. 
	
	Considering equations (50) ,(54) and approximate relation for the commutator $ P_1 L = L P_1 - [P_{1}, L] $ one obtains:
	\begin{equation}\label{pir}
	  \Pi_{t}= \sqrt{bc} \Pi_{x}, 
	 	\end{equation}
\begin{equation}\label{lar}
	  \Lambda_{t}=-\sqrt{bc}\Lambda_{x}.
 	\end{equation}
	 Solving the first order equations by method of characteristics give $u, v$ by \eqref{uv}; formally the system coincides with \eqref{Pi3} but velocity of propagation and coefficients in \eqref{uv} are  functions depending on coordinate and.
		
		\section{Acoustics example}
		
The linear version of the momentum (Euler) linear equation for a compressible liquid is
 \begin{equation}\label{EulerL}
		{\rho}\frac{\partial \vec v}{\partial t}=- \nabla p.
		\end{equation}
	The continuity equation reads		
			\begin{equation}\label{EulerL0}
 \frac{\partial \rho}{\partial t}+ \nabla(\rho \vec v)=0
\end{equation}
and, together with equations  of state close a dynamic problem for a fluid perturbations.
 
Consider   a linearization in one-dimensional case via the perturbations marked by primes $\rho=\rho_0(x)+\rho'(x,t)$, $p=p_0(x)+p'$ , $v=v'$
\begin{equation}\label{EulerL1}
\rho_0\frac{\partial v'}{\partial t}=-\frac{\partial p'}{\partial x}, 
		\end{equation}
\begin{equation}\label{EulerL2}
\frac{\partial\rho'} {\partial t}+\rho_0 \frac{\partial v'  }{\partial x}+ v^{'} \frac{\partial\rho_0}{\partial x}=0,
 \end{equation}
It is known that such case without dissipation leads to adiabatic condition:
 \begin{equation}
\frac{p}{\rho^\gamma}=\frac{p_0}{\rho_0^\gamma}
  \end{equation}
Its account results in the system: 
	\begin{equation}
(\rho_0 v {'})_t + \frac{p_0}{\rho_0} \gamma \rho'_x = 0,
\end{equation}
\begin{equation} 
\rho'_t + (\rho_0 v')_x = 0.
\end{equation}

Notations $u = \rho$, $\rho_0 v{'} =v $ and $b = 1$ , $ \frac{p_0}{\rho_0} \gamma = c(x)$ establish   this system correspondence with the Eqs. (\ref{1},\ref{2}).
In this case $ f(x) = \sqrt{ \gamma\frac{p_0}{\rho_0}} $ is the propagation velocity  of acoustic wave. So, the combination of the pressure and velocity perturbations as in the relations for projecting operators that define the right and left waves \eqref{pi},\eqref{la} solve the problem of such wave initialization. The evolution in the weakly inhomogeneous medium in the first approximation is solved by characteristic method mentioned in connection with \eqref{pir},\eqref{lar}.

\section{Conclusion}
	Development of the method of dynamic projection operators for the theory of hyperbolic systems of partial differential equations with variable coefficients was considered and results in the separated first order system (\ref{pir},\ref{lar}). this result is obtained in lowest approximation. next approximation give more exact description, but the procedure principally the same. The statement of problem for the separated equations and reverse transformation to original variables is realized by means of the projecting operators as in \eqref{pi} and its inverse \eqref{uv}.  This idea to built approximate solutions  for such systems  was presented in \cite{Jacobi}.	A nonlinearity   may be introduced as a perturbation by amplitude parameter, application of the projecting leads to the modes interaction.	Taking the elliptic case one concludes that it could be applied for example to boundary problem of Laplace/Poisson equation at half-plane.

\end{document}